\documentclass[fleqn,twoside]{article}
\usepackage{espcrc2}
\usepackage[T1]{fontenc}
\usepackage[latin1]{inputenc}
\usepackage{graphicx}
\usepackage{epsfig}
\usepackage{amssymb}
\usepackage{amsmath}
\usepackage{latexsym}
\newcommand{\EJP}{{\it Eur. J. Phys.} } 

\newcommand{\jpg}{{\it J. Phys. G: Nucl. Part. Phys.} }   

\newcommand{\NP}{{\it Nucl. Phys.} }

\newcommand{\PRL}{{\it Phys. Rev. Lett.} }

\newcommand{\etal}{{et al.\/}}

\title{The SPL-Fréjus physics potential}

\author{Jean-Eric Campagne\address{Laboratoire de l'Accélérateur Linéaire -
Univ. Paris-Sud - CNRS - BP 34 -
91898 Orsay Cedex, France}}

\begin{document}
\begin{abstract}
An optimization of the CERN-SPL beam line has been performed which leads to better sensitivities to the $\theta_{13}$ mixing angle and to the $\delta_{CP}$ violating phase than those advocated considering baseline scenario. 
\end{abstract}

\maketitle

\section{Introduction}
An optimization of the energy as well as the secondary particle focusing and decay tunnel has been undertaken in the context of a Super Beam of 4~MW \cite{JECACLAL} using the CERN-SPL \cite{SPL} and searching for $\nu_\mu \rightarrow \nu_e$ ($\bar{\nu}_\mu \rightarrow \bar{\nu}_e$) appearance channels in a 440~kT fiducial volume water Cerenkov detector located in an possible new underground laboratory at the Fréjus tunnel, 130~km from the CERN complex.
\section{Fluxes}
The secondary particles from the interaction of proton beam impinging a 30~cm long 1.5~cm diameter mercury target, have been obtained with the FLUKA generator. At kinetic energy of 3.5 (2.2)~GeV, the number of p.o.t per year is 0.69 (1.10) $10^{23}$ while the numbers of $\pi^+/\pi^-/K^+/K^o$ per p.o.t are $0.41/0.37/35 10^{-4}/30 10^{-4}$ ($0.24/0.18/7 10^{-4}/6 10^{-4}$). At higher beam energy, the kaon rates grow rapidly compared to the pion rates,  and needless to emphasize the need of an experimental confirmation \cite{HARP,MINERVA} of such numbers.

\begin{figure}
\centering
\includegraphics[height=60mm]{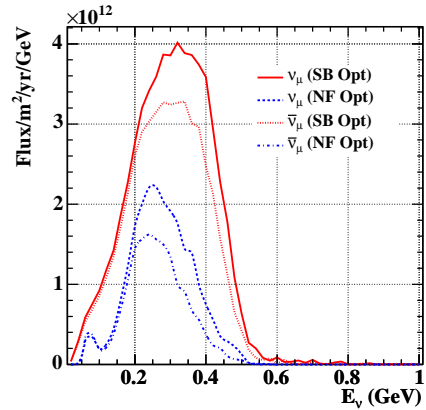}
\caption{\label{fig:fluxComparison}The $\nu_\mu$ and the $\bar{\nu}_\mu$ fluxes obtained by optimizing the SPL for a Super Beam case ("SB Opt.") are compared to those obtained with a focusing system designed for a Neutrino Factory ("NF Opt.").}
\end{figure}

The focusing system optimized in the context of a Neutrino Factory \cite{SIMONE1,DONEGA} has been redesigned considering the specific requirements of a Super Beam. The most important points are the phase spaces that are covered by the two types of horns are different, and that the pions to be focused should have an energy of the order of $p_\pi (\mathrm{MeV})/3 \approx E_\nu \gtrsim  2L(\mathrm{km})$ to obtain a maximum oscillation probability. In practice, this means that one should collect $800$~MeV/c pions.
%
%
The resulting fluxes for the positive ($\nu_\mu$ beam) and the negative focusing ($\bar{\nu}_\mu$ beam) are show on figure \ref{fig:fluxComparison}. The total number of $\nu_\mu$ ($\bar{\nu}_\mu$) in positive (negative) focusing is about $1.18 (0.97) 10^{12}/\mathrm{m}^2/\mathrm{yr}$ with an average energy of $300$~MeV. The $\nu_e$ ($\bar{\nu}_\mu$) contamination in the $\nu_\mu$ beam is around $0.7\%$ ($6.0\%$). Compared to  the fluxes used in references \cite{MEZZETTONF02,DONINI04} the gain is at least a factor $2.5$. Using neutrino cross-sections on water \cite{LIPARIxsec}, the number of expected $\nu_\mu$ charged current is about $95$ per kT.yr.
\section{Physics potential}
The physics potential of the new optimized SPL may be determined using GLoBES software \cite{GLOBES}. Both the appearance and the disappearance channels have been used, and also five bins of 200~MeV each have been introduced. The $\pi^o$ background have been rejected using a tighter PID cut compared to standard SuperK analysis \cite{MEZZETTONF02}. The Michel electron has been required for the $\mu$ identification. As ultimate goal suggested by \cite{T2K} a 2\% systematical error is used both for signal and background.

\begin{figure}
\centering
\includegraphics[height=60mm]{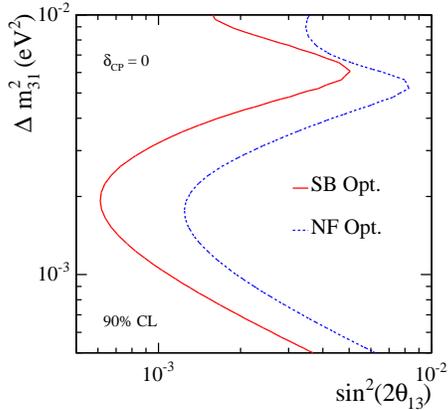}
\caption{\label{fig:SPLDmTheta13Comparison}Comparison of the sensitivity contours with the new optimized setup and the original SPL design after 5 years of running with positive focusing.}
\end{figure}

\begin{figure}
\centering
\includegraphics[height=60mm]{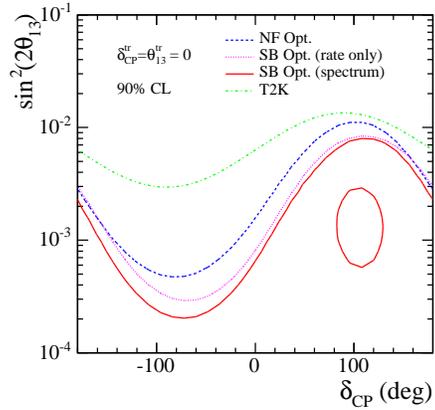}
\caption{\label{fig:SPLCPSensi}Comparison of the sensitivity on combined $\theta_{13}$ and $\delta_{CP}$ after 5 years of positive focusing. The T2K sensitivity contour has been derived from  reference \cite{T2K}. No mass hierarchy nor octant hierarchy ambiguity has been considered.}
\end{figure}

The $90\%$ CL sensitivity contour in the ($\Delta m^2_{31}$,$\sin^22\theta_{13}$) plane after 5 years running with a positive focusing is shown on figure \ref{fig:SPLDmTheta13Comparison}. With the new optimized setup, one expects to reach a limit on $\theta_{13}$ of $0.7^o$.
One may also appreciate on figure \ref{fig:SPLCPSensi} the improvement on the combined sensitivity of $\sin^22\theta_{13}$ and $\delta_{CP}$ with the new optimization compared for instance to the T2K project \cite{T2K}. 

So, the optimized SPL beam line operating a Super Beam towards a megaton scale detector at the Fréjus tunnel has a great potential contrary to the strong statement advocated at the last SPSC/PSC Villars meeting \cite{VILLARS}. 
\end{document}